\documentclass[]{aastex62}
\received{\today}
\revised{}
\accepted{}
\submitjournal{ApJL}

\shorttitle{M31 spectrum}
\shortauthors{Battistelli et al.}

\begin{document}

\title{Strong evidence of Anomalous Microwave Emission from the flux density spectrum of M31}

\correspondingauthor{Elia Stefano Battistelli, Sofia Fatigoni, Matteo Murgia}
\email{elia.battistelli@roma1.infn.it, sfatigoni@phas.ubc.ca, matteo.murgia@inaf.it}

\author[0000-0001-5210-7625]{E.S. Battistelli}
\affil{Sapienza - University of Rome - Physics department, Piazzale Aldo Moro 5 - I-00185, Rome, Italy}
\affil{INAF - IAPS, Via Fosso de Cavaliere, 100 - I-00133, Rome, Italy}
\affiliation{INFN - Sezione di Roma, Piazzale Aldo Moro 5 - I-00185, Rome, Italy}

\author{S. Fatigoni}
\affiliation{University of British Columbia, Department of Physics and Astronomy, Vancouver, BC, V6T 1Z1, Canada}

\author[0000-0002-4800-0806]{M. Murgia}
\affiliation{INAF - Osservatorio Astronomico di Cagliari, Via della Scienza 5 - I-09047 Selargius (CA), Italy}

\author{A. Buzzelli}
\affil{Sapienza - University of Rome - Physics department, Piazzale Aldo Moro 5 - I-00185, Rome, Italy}
\affiliation{INFN - Sezione di Roma II, Via della Ricerca Scientifica, 1 - I-00133, Rome, Italy}

\author[0000-0002-3973-8403]{E. Carretti}
\affiliation{INAF - Istituto di Radioastronomia - Via P. Gobetti, 101 -  I-40129 Bologna, Italy}

\author{P. Castangia}
\affiliation{INAF - Osservatorio Astronomico di Cagliari, Via della Scienza 5 - I-09047 Selargius (CA), Italy}

\author{R. Concu}
\affiliation{INAF - Osservatorio Astronomico di Cagliari, Via della Scienza 5 - I-09047 Selargius (CA), Italy}

\author{A. Cruciani}
\affiliation{INFN - Sezione di Roma, Piazzale Aldo Moro 5 - I-00185, Rome, Italy}

\author[0000-0001-6547-6446]{P. de Bernardis}
\affil{Sapienza - University of Rome - Physics department, Piazzale Aldo Moro 5 - I-00185, Rome, Italy}
\affiliation{INFN - Sezione di Roma, Piazzale Aldo Moro 5 - I-00185, Rome, Italy}

\author[0000-0001-5479-0034]{R. Genova-Santos}
\affiliation{Instituto de Astrofisica de Canarias, C/Via Lactea s/n, E-38205 La Laguna, Tenerife, Spain}
\affiliation{Departamento de Astrofísica, Universidad de La Laguna (ULL), E-38206 La Laguna, Tenerife, Spain}

\author[0000-0003-3644-3084]{F. Govoni}
\affiliation{INAF - Osservatorio Astronomico di Cagliari, Via della Scienza 
5 - I-09047 Selargius (CA), Italy}

\author{F. Guidi}
\affiliation{Instituto de Astrofisica de Canarias, C/Via Lactea s/n, E-38205 La Laguna, Tenerife, Spain}
\affiliation{Departamento de Astrofísica, Universidad de La Laguna (ULL), E-38206 La Laguna, Tenerife, Spain}

\author{L. Lamagna}
\affil{Sapienza - University of Rome - Physics department, Piazzale Aldo Moro 5 - I-00185, Rome, Italy}
\affiliation{INFN - Sezione di Roma, Piazzale Aldo Moro 5 - I-00185, Rome, Italy}

\author{G. Luzzi}
\affil{Sapienza - University of Rome - Physics department, Piazzale Aldo Moro 5 - I-00185, Rome, Italy}
\affil{Italian Space Agency - Via del Politecnico snc  - I-00133, Rome, Italy }

\author[0000-0001-5105-1439]{S. Masi}
\affil{Sapienza - University of Rome - Physics department, Piazzale Aldo Moro 5 - I-00185, Rome, Italy}
\affiliation{INFN - Sezione di Roma, Piazzale Aldo Moro 5 - I-00185, Rome, Italy}

\author{A. Melis}
\affiliation{INAF - Osservatorio Astronomico di Cagliari, Via della Scienza 5 - I-09047 Selargius (CA), Italy}

\author{R. Paladini}
\affiliation{Infrared Processing Analysis Center, California Institute of Technology, Pasadena, CA 91125, USA}

\author[0000-0002-5444-9327]{F. Piacentini}
\affil{Sapienza - University of Rome - Physics department, Piazzale Aldo Moro 5 - I-00185, Rome, Italy}
\affiliation{INFN - Sezione di Roma, Piazzale Aldo Moro 5 - I-00185, Rome, Italy}

\author{S. Poppi}
\affiliation{INAF - Osservatorio Astronomico di Cagliari, Via della Scienza 5 - I-09047 Selargius (CA), Italy}

\author{F. Radiconi}
\affil{Sapienza - University of Rome - Physics department, Piazzale Aldo Moro 5 - I-00185, Rome, Italy}

\author[0000-0003-3767-7085]{R. Rebolo}
\affiliation{Instituto de Astrofisica de Canarias, C/Via Lactea s/n, E-38205 La Laguna, Tenerife, Spain}
\affiliation{Departamento de Astrofísica, Universidad de La Laguna (ULL), E-38206 La Laguna, Tenerife, Spain}

\author[0000-0001-5289-3021]{J.A. Rubino-Martin}
\affiliation{Instituto de Astrofisica de Canarias, C/Via Lactea s/n, E-38205 La Laguna, Tenerife, Spain}
\affiliation{Departamento de Astrofísica, Universidad de La Laguna (ULL), E-38206 La Laguna, Tenerife, Spain}

\author{A. Tarchi}
\affiliation{INAF - Osservatorio Astronomico di Cagliari, Via della Scienza 5 - I-09047 Selargius (CA), Italy}

\author{V. Vacca}
\affiliation{INAF - Osservatorio Astronomico di Cagliari, Via della Scienza 5 - I-09047 Selargius (CA), Italy}



\begin{abstract}

We have observed the Andromeda galaxy, Messier 31 (M31), at 6.7~GHz with the Sardinia Radio Telescope. We mapped the radio emission in the $C$-band, re-analyzed $WMAP$ and $Planck$ maps, as well as other ancillary data, and we have derived an overall integrated flux density spectrum from the radio to the infrared. This allowed us to estimate the emission budget from M31. Integrating over the whole galaxy, we found strong and highly significant evidence for anomalous microwave emission (AME), at the level of 1.45$^{+0.17}_{-0.19}$~Jy at the peaking frequency of $\simeq$25~GHz. Decomposing the spectrum into known emission mechanisms such as free-free, synchrotron, thermal dust, and AME arising from electric dipole emission from rapidly rotating dust grains, we found that the overall emission from M31 is dominated, at frequencies below 10~GHz, by synchrotron emission with a spectral index of -1.10$^{+0.10}_{-0.08}$, with subdominant free-free emission. At frequencies $\gtrsim$~10~GHz, AME has a similar intensity to that of synchrotron and free-free emission, overtaking them between 20~GHz and 50~GHz, whereas thermal dust emission dominates the emission budget at frequencies above 60~GHz, as expected. 

\end{abstract}



\section{Introduction} \label{sec:intro}

Messier 31 (M31) is the largest and most massive galaxy in the Local Group. Thanks to its proximity and similarity to our own Galaxy, Andromeda represents a unique laboratory in which to study effects and emissions arising from extragalactic sources and to understand the astrophysics of our own Galaxy seen from an external point of view.

M31 is a widely studied astrophysical object at all wavelengths, and observations have been undertaken in several bands from gamma-ray to radio wavelengths. Of interest for the present study is the analysis by \cite{fri12}, who used infrared (IR) data obtained from the \textit{Herschel} satellite at five wavelengths from 100 to 500~$\mu$m. The millimetric and centimetric emission from M31 is well summarized by \cite{pla15} (hereafter PLA15). Radio emission of M31 is also quite well studied (e.g., \citealp{has81, gal12}), and 5~GHz continuum observations over a field of 2$^{\circ}$.5$\times$1$^{\circ}$.7 have been carried out with the Effelsberg telescope \citep{ber03}, focusing mainly on polarized synchrotron emission from the disk and its magnetic field. Interestingly, in the microwave range, aside from the low-angular-resolution maps of $WMAP$ and $Planck$ Low Frequency Instrument (LFI), and the Effelsberg measurements of the disk at 5~GHz, there is a gap in the observations.  This is, for instance, described by PLA15, who reported the aperture photometry analysis obtained integrating over the whole galaxy.

The emission budget from astrophysical sources at microwave frequencies is mostly dominated by the well-studied and well-understood free-free, synchrotron, and thermal dust emission. Nevertheless, observations mainly carried out in our Galaxy have revealed an unexpected excess of emission in the microwave band from 10 to 50~GHz that cannot be explained by standard emission mechanisms or in terms of the cosmic microwave background (CMB). This excess emission (anomalous microwave emission, AME), first observed in the 1990s as dust-correlated emission in COBE maps \citep{kog96}, is observable both as diffuse emission and in selected sky regions (see \citealp{dick18} for a recent review). Its physical origin is not fully understood yet, but the most convincing models predict that AME is dominated by electric dipole emission from rapidly rotating dust grains (spinning dust; \citealp{dra98}). Other physical emission mechanisms, such as hot free-free, hard synchrotron, or magnetic dipole emission, however, should play an important role in the AME budget and can be disentangled from spinning dust owing to their spectral behavior in the 6-25~GHz band and polarization \citep{dick18}. Also, high angular resolution studies highlight the complexity of the emission of some Galactic regions with far from explained phenomenology \citep{bat15,pal15,cru16,hen17}. The importance of a full understanding of AME depends not only on our comprehension of the astrophysical mechanisms at its origin, but also on the need for CMB experiments in order to understand and remove foreground signals (e.g., \citealp{bic15}). 

Some Galactic regions have been quite well studied and characterized (e.g., \citealp{wat05,bat06,gen15}). Typically, AME is found near HII regions, molecular clouds, and possibly supernovae remnants (e.g. \citealp{pla14b}). However, in a recent study, AME was found in three protoplanetary disks, the only known systems hosting hydrogenated nano-diamonds \citep{gre18}. Of great interest is clearly the possibility of detecting AME from extragalactic sources as this would represent a unique possibility to study astrophysical processes mainly studied only in our Galaxy. 

Extragalactic evidence of AME has been found in a limited number of cases. \cite{mur10} reported the first extragalactic evidence of AME in selected star-forming regions in the spiral galaxy NGC~6946 with the Green Bank Telescope. Follow-up observations confirmed the excess \citep{sca10, hen15}. \cite{pla11} reported emission from the Small Magellanic Cloud that was partly interpreted as spinning dust. The interpretation is however complicated by additional emission from thermal dust with possible contamination from magnetic dipole emission \citep{dra12}. Using the Very Large Array (VLA), \cite{mur18} found evidence of AME in the compact radio source NGC~4725~B, located 1.9  kpc from the nucleus of NGC~4725. A tentative 2.3~$\sigma$ detection of AME in M31 has been reported by PLA15, who integrated the emission over the whole galaxy. These detections represent a unique step forward in the comprehension of AME. Nevertheless, none of the above observations have been able to characterize the microwave emission arising from an entire galaxy with sufficient sensitivity to provide an unambiguous explanation. 

\section{Observations and data analysis} \label{sec:obs}

The integrated flux densities from M31 presented in this Letter are part of the observations undertaken in 2016 during the Early Science Commissioning Phase of the INAF-Sardinia Radio Telescope (SRT) \citep{bol15,mur16,pra17}. With its 64~m primary mirror, the SRT, at 6.7~GHz, has a beam primary lobe width at half maximum (FWHM) of $\simeq$2$'$.9. Among the backends currently available, we used the SArdinia Roach2-based Digital Architecture for Radio Astronomy (SARDARA) backend \citep{mel18}, a wide band digital backend, based on the ROACH2\footnote{https://casper.berkeley.edu/wiki/ROACH2} technology, which can divide the signal in the given bandwidth into up to 16,384 channels for full Stokes spectropolarimetric observations. 

We mapped M31 over an RA$\times$Dec=2$^{\circ}$.4$\times$3$^{\circ}$.1= 7.4~deg$^{2}$ rectangle centred on the IR core of M31 at (R.A.;~decl.) = (0h~42m~48s; +41$^{\circ}$~16$'$ ~48$''$; J2000). The dimensions of this map have been selected to provide the minimum contour that includes the galaxy, to which we added a surrounding map edge of 0$^{\circ}$.25, in each of the two directions, estimated by comparing the two maps provided by \textit{Herschel}-SPIRE at 250~$\mu$m \citep{fri12} and Planck at 857~GHz \citep{pla14a}, which trace the galaxy in its maximum extension in terms of dust content. During observations we adopted an on-the-fly map scanning strategy, 6~$'$/s orthogonal sub-scans along the R.A. and decl directions spaced by 54$''$. Mapping the whole area required 209 R.A. sub-scans and 161 decl. sub-scans. The observations were completed in 64~hr, during which we carried out 44 complete scans of the galaxy. 

Data analysis was carried out using the Single-dish Spectral-polarimetry Software (SCUBE, see \citealp{mur16}), proprietary C++ software for calibrating, imaging, and analyzing data acquired with the SRT. The spectral capability of our observations allowed us to clean data for radio frequency interference (RFI) contamination and corrupted scans.  Data reduction included bandpass, flux density, and polarization calibration. The calibration sources used were 3C84, 3C48, 3C138, 3C147, NGC~7027, 3C286, and 3C295, with flux densities derived according to the scale of \cite{per13}. The overall flux density calibration was performed using the quasar 3C147, and the systematic uncertainty is estimated to be 5\%. The other calibrators were used for polarization and consistency checks.

Before stacking the sub-scans to form the maps, each sub-scan underwent a linear baseline removal procedure with the aim of removing unwanted foreground signals and receiver instability. We removed the baseline from each M31 sub-scan individually using an iterative approach. As a first step, a mask was used to exclude the regions occupied by M31 itself and by the strongest point-like sources in the field. These ``cold sky'' regions were modeled using a linear least-squares fit. In the refinement steps, a model of the sky emission was formed by stacking all baseline-subtracted sub-scans. We then performed a new baseline removal from the individual sub-scans by fitting the best second-order polynomial parameters that minimized the difference between the data and the model. We applied an automated flagging procedure to remove isolated low-level RFI, and then we stacked all the baseline-subtracted scans to obtain a new higher signal-to-noise level model from the final map. 

Given our on-the-fly scanning strategy, the final map still presents typical features along the direction of the single sub-scans because of poorly removed RFI or short-time fluctuations both in the atmospheric opacity and in the receiver gain. This noise was not totally removed during the flagging and the baseline removal process, and it can persist in the final map in the form of stripes oriented along the scanning directions. However, these features can be isolated and removed by analyzing maps acquired in different directions (e.g. \citealp{eme88}). SCUBE implements a wavelet method to stack a set of maps taken along orthogonal directions (see \citealp{mur16}). In the creation of the final maps, we combined all scans by noise weighted-averaging them to create the final intensity maps.

Our maps are possibly still characterized by foreground signal residuals and confusion noise from unresolved radio sources. In order to extract the emission information arising only from M31, we masked out an elliptical region with a major axis of 91$'$.5 and minor axis of 59$'$.5 (position angle, PA=-52$\deg$, East-to-North), centered on the galaxy core and we reconstructed the surrounding base level by applying the Papoulis-Gerchberg Algorithm, PGA  \citep{pap75}. The PGA is a popular technique that can be used to reconstruct missing data in band-limited signals. The reconstructed base-level image is subtracted from the final map allowing the removal of features connected to the baseline removal and map-making, as well as foreground/background emission reducing the map to zero level outside the galaxy.

\begin{figure}
\plotone{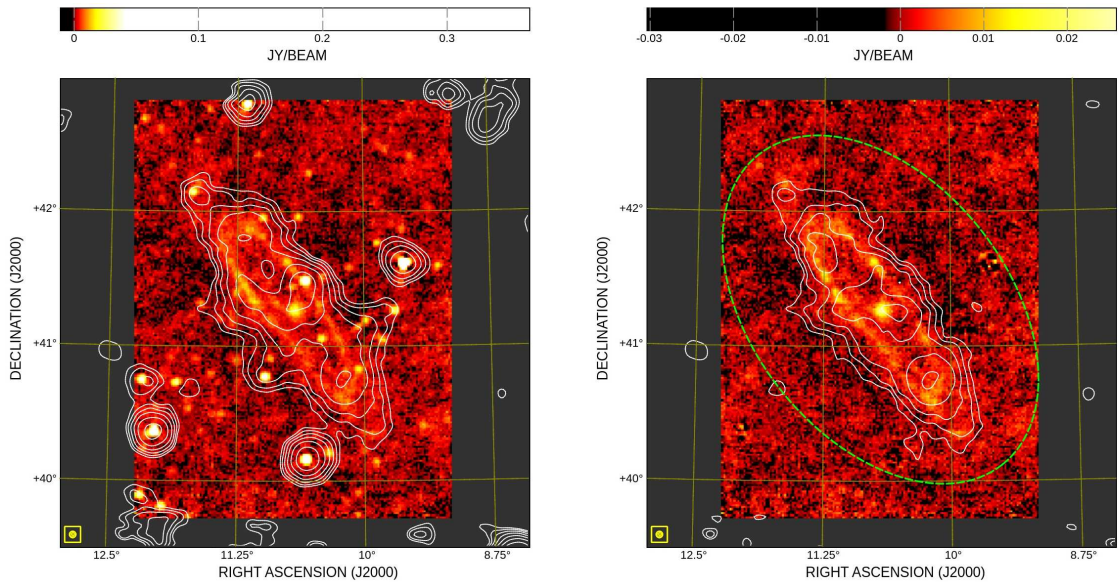}
\caption{Left panel: continuum map obtained at 6.7~GHz with the SRT. White contours refer to the SRT image at 1.4~GHz \citep{mel18} (levels start at 90 mJy/beam and scale by $\sqrt{2}$). Right panel: the same map is shown after removing the point sources. Also indicated is the elliptical area over which the aperture photometry was performed. In the bottom-left corner the 2$'$.9 FWHM beam is indicated. \label{fig:map}}
\end{figure}

In Fig.~\ref{fig:map} we present the final $C$-band continuum map obtained by combining all the spectral channels within the effective 1.25~GHz bandwidth. The  presence of several point-like radio sources is evident. Using the NRAO VLA Sky Survey (NVSS) map at 1.4GHz \citep{con98} and other ancillary low-frequency maps, we identified  $\simeq$600 radio sources down to 1 mJy flux density level in our $C$-band map. A detailed analysis of the $C$-band continuum map and sources therein, as well as of the morphology, will be presented elsewhere (E. S. Battistelli et al. 2019, in preparation). In the source-subtracted map (note that point sources were subtracted before the PGA was applied; see Fig.~\ref{fig:map} and Sect. \ref{sec:ancillary} for details), we calculated a final rms value of $S_{\rm rms}$=0.49mJy/beam, consistent with the expected confusion noise level: $S_{\rm conf}$= 0.44 mJy/beam \citep{con74, dez18}. The same analysis was  applied to two different maps obtained by dividing the overall spectral bandwidth into two sub-bands, each with a $\simeq$~625~MHz bandwidth.

\section{Aperture photometry on $C$-band and ancillary data}\label{sec:ancillary}

We analyzed a set of different maps ranging from radio to IR frequencies in order to extract the integrated flux density from M31 to be compared with the $C$-band observations presented in this work. At frequencies below the $C$-band, we used $L$-band measurements at 1.4~GHz obtained with the SRT during SARDARA commissioning \citep{mel18}. In order to monitor only the continuum radio emission from Andromeda, we filtered out the 1.4~GHz  signal from neutral hydrogen arising from both our Galaxy and M31, excluding the frequency range 1420.65~MHz-1423.9~MHz. Data at 34.5~MHz \citep{dwa90}, 408~MHz \citep{has81}, and 1.4~GHz (\citealp{rei82} for confirmation of the SRT $L$-band flux densities) were used to cover the frequency range where synchrotron and free-free are expected to be the dominant sources of emission. Unfortunately, given the aperture photometry that we are applying, we were unable to use the 5~GHz continuum observations carried out with the Effelsberg telescope \citep{ber03} because of the limited extension of the map. Above the $C$-band, we analyzed $WMAP$ 9-year data \citep{ben13}, $Planck$ LFI and HFI (DR2, \citealp{pla15}) CMB subtracted maps (i.e., SMICA\footnote{http://pla.esac.esa.int/pla}), as well as \textit{Herschel} maps obtained within the HELGA project \citep{fri12}.  

\begin{figure}
\plotone{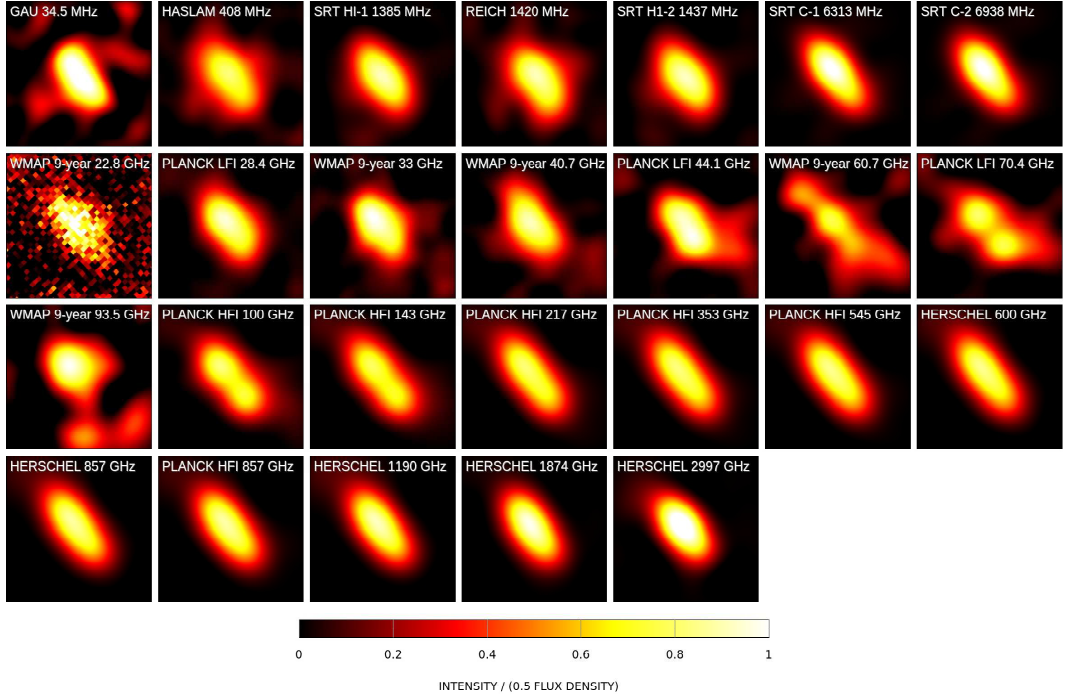}
\caption{Convolved and background-subtracted maps over which we performed the aperture photometry analysis. Each map was normalized to its overall emission reported in Table \ref{obs} with color scale ranging from zero to half the total flux density. \label{fig:mapall}}
\end{figure}

All of the maps were subjected to consistent analysis based on the $C$-band scanning strategy, reprocessing, and map dimensions. We carried out the same aperture photometry analysis in order to extract the total flux densities. For self-consistency in the flux density estimation arising only from M31, point sources showing a signal 3 $\sigma$ above the noise estimation were subtracted from our maps and from the ancillary maps. Using the same approach applied to the $C$-band maps, point sources outside M31 were identified using the 1.4~GHz NVSS survey \citep{con98} and fitted in frequency with second-order polynomial fits using Cats extragalactic database\footnote{https://www.sao.ru/cats/} \citep{ver97}. This was applied to all the detected sources in each map with the exception of the BLAZAR B3-0035+413, owing to its extreme intensity and variability, for which we simply fitted its emission by assuming, as prior, the coordinates of the center position, as well as the instrumental beam. This method allowed us to calculate and extrapolate point sources for each frequency and subtract them from all the maps we were using. After removing point sources, in order to isolate the emission from the galaxy itself from foreground and background emission, the PGA was applied to all the maps. Finally, we convolved every map to the lowest angular resolution (i.e., 51$'$.3 of the $WMAP$ 22.8~GHz channel). The reprocessed and convolved maps are shown in Figure \ref{fig:mapall}. 

The final flux densities were obtained by integrating over the same elliptical region of 91$'$.5 and 59$'$.5 axes (effective radius 73$'$.8). The background was estimated over the surrounding region filling the entire $C$-band map. It is worth noting that the PGA actually evaluates the background over the entire map (with the exception of the masked region) and we then remove it from the map. In this sense, applying the PGA can be seen as an extension to higher Fourier modes of the simple removal of an average background over an annular region around the source. The background removal is thus redundant in this analysis. 

The stability of the choice of the convolving resolution, as well as the background subtraction strategy, was carefully monitored. We found that flux densities, obtained with a simple background removal arising from the average signal outside the mask, give consistent results. Also, convolving our maps at angular resolutions up to 51$'$.3 does not affect the measured flux densities by more than 10\%, which includes both the flux density leakage from the integration ellipse to the surrounding region and possible contamination of out-of-map regions for those maps characterized by the lowest angular resolution. Also, results were monitored by changing the integrating ellipse size, ellipticity, and inclination angle. These analyses showed that the actual chosen values are the most appropriate with deviations lower than 5\%, with the ellipse inclination angle ranging between $-22^{\circ}$ and $-81^{\circ}$, eccentricity as low as 0.53 (fixing the major axis), and within effective radius down to 67'. For the $C$-band SRT maps, this resulted in overall flux densities of 1.207$\pm$0.084~Jy and 1.191$\pm$0.089~Jy, respectively, at 6.3125 and 6.9375\,GHz, where uncertainties reflect only the statistical fluctuations. The aforementioned systematic check was considered in the spectrum fit (see Sec.\ \ref{sec:sed}), accounting for an additional 12\% systematic uncertainty to all flux densities arising from the flux density extraction itself. The results are summarized in Table \ref{obs}.

\begin{deluxetable}{lccccl}
\tablecaption{Flux densities from M31 calculated on the $C$-band SRT Maps and Ancillary Data. \label{obs}}
\tablehead{
\colhead{Map} & \colhead{Frequency (GHz)} & \colhead{FWHM ($'$)} & \colhead{Flux Density (Jy)}  & \colhead{Calibration Unc.} & \colhead{References}
}
\startdata
GAU$_{34.5}$ & 0.0345 & 48 &92$\pm$17  &10\%& \cite{dwa90} \\
Haslam & 0.408 & 51 & 18.4$\pm$1.6  &5\%& \cite{has81} \\
SRT HI-1 & 1.385 & 13.94 & 5.43$\pm$0.41  &5\%&  \cite{mel18} \\
Reich & 1.420 & 35.4 & 5.28$\pm$0.41  &10\%& \cite{rei82} \\
SRT HI-2 & 1.437 & 13.66 & 5.27$\pm$0.38  &5\%&  \cite{mel18} \\
SRT C-1 & 6.313 & 2.897 & 1.207$\pm$0.084  &5\%& This work \\
SRT C-2 & 6.938 & 2.736 & 1.191$\pm$0.089  &5\%&  This work \\
$WMAP$ 9-year & 22.8 & 51.3 & 2.00$\pm$0.17  &3\%& \cite{ben13} \\
$Planck$ LFI & 28.4 &  33.1& 1.86$\pm$0.15  &3\%& PLA15 \\
$WMAP$ 9-year & 33.0 & 39.1 & 1.71$\pm$0.21  &3\%&  \cite{ben13} \\
$WMAP$ 9-year & 40.7 & 30.8 & 1.31$\pm$0.16  &3\%&  \cite{ben13} \\
$Planck$ LFI & 44.1 & 27.9 & 1.45$\pm$0.25  &3\%& PLA15 \\
$WMAP$ 9-year & 60.7 & 21.0 & 1.72$\pm$0.42  &3\%&  \cite{ben13} \\
Planck LFI & 70.4 & 13.1 & 2.12$\pm$0.36  &3\%& PLA15 \\
$WMAP$ 9-year & 93.5 & 14.8 & 3.5$\pm$1.0  &3\%&  \cite{ben13} \\
$Planck$ HFI & 100 & 9.65 & 5.78$\pm$0.53  &3\%& PLA15 \\
$Planck$ HFI &143 & 7.25 & 15.7$\pm$1.4  &3\%& PLA15 \\
$Planck$ HFI & 217 & 4.99 & 69.4$\pm$5.5  &3\%& PLA15 \\
$Planck$ HFI & 353 & 4.82 & 318$\pm$24  &3\%& PLA15 \\
$Planck$ HFI & 545 & 4.68 & 1027$\pm$73  &7\%& PLA15 \\
\textit{Herschel} & 600 & 0.59 & 1195$\pm$85  &4\%& \cite{fri12} \\
\textit{Herschel} & 857 & 0.4 & 2830$\pm$180  &4\%& \cite{fri12} \\
Planck HFI & 857 & 4.33 & 3020$\pm$190  &7\%& PLA15 \\
\textit{Herschel} & 1199 & 0.29 & 5330$\pm$370  &4\%& \cite{fri12} \\
\textit{Herschel} & 1874 & 0.22 & 7020$\pm$230  &5\%& \cite{fri12} \\
\textit{Herschel} & 2997 & 0.21 & 2980$\pm$140  &5\%& \cite{fri12} \\
\enddata
\end{deluxetable}

\begin{figure}
\plotone{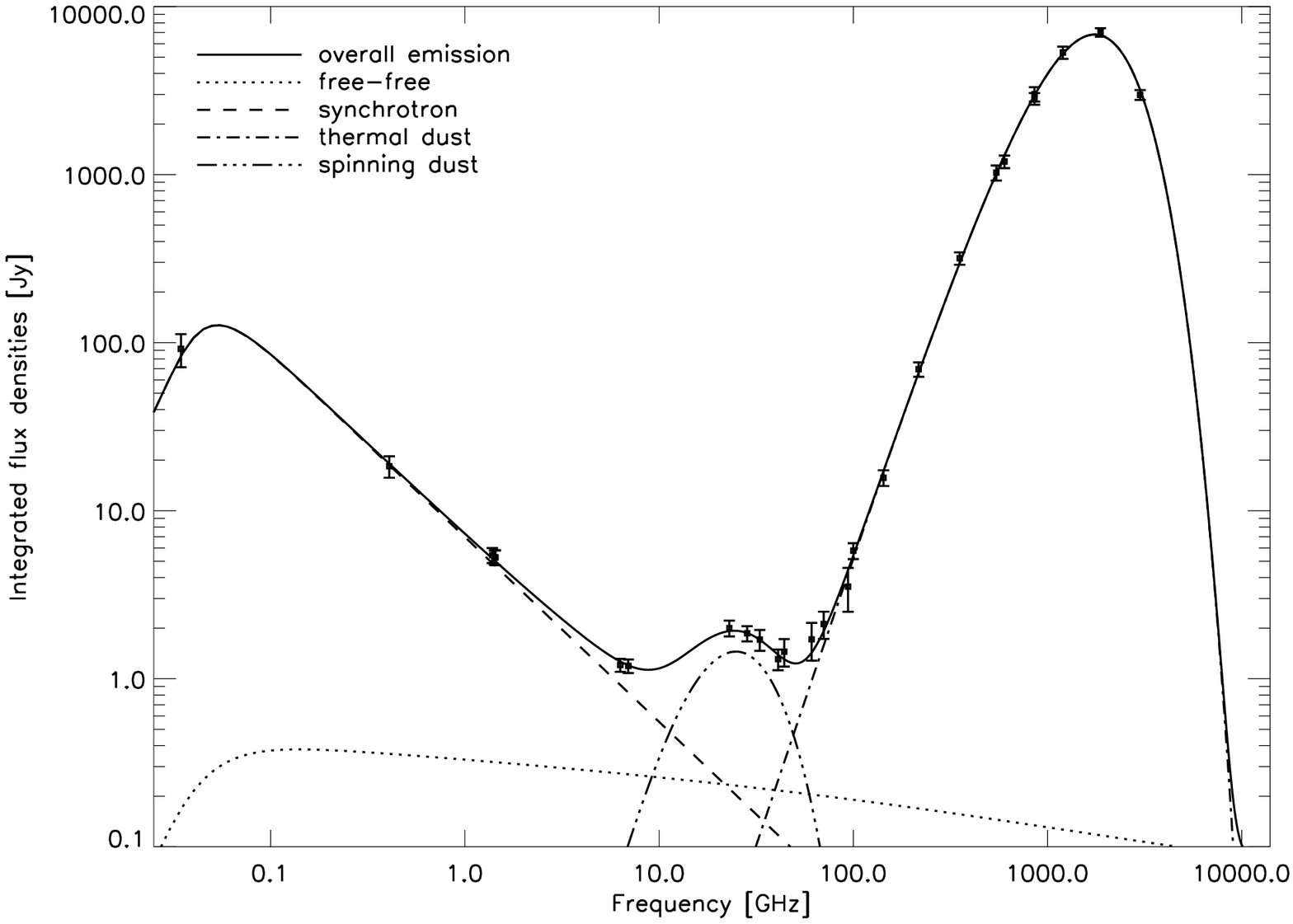}
\caption{Flux density spectrum and best fit arising from aperture photometry over M31 (see the text for details). \label{fig:sed}}
\end{figure}

\section{Galaxy spectrum and data fit} \label{sec:sed}

In order to extract astrophysical information about the emission budget of Andromeda, we calculated the overall spectrum and fitted it over known continuum emissions from the radio to IR wavelength range  through a superposition of different kinds of emission using a Python implementation of the Goodman and Wear's Markov chain Monte Carlo (MCMC) Ensemble sampler \citep{goo10, formac13}. 
We present the spectrum obtained from the entire galaxy of M31 in Figure \ref{fig:sed}. For the synchrotron emission, $S_{\rm sy}$, we assumed a dependence $S_{\rm sy}=A_{\rm sy}\nu^{\alpha}$, leaving the spectral index $\alpha$ and the amplitude $A_{\rm sy}$ as free parameters, while for the free-free emission we fitted the amplitude $A_{\rm ff}$ at 1~GHz, assuming the spectral behavior following PLA15 with a Gaunt factor determined as in \cite{dra11}. For both thermal and nonthermal emission we accounted for an average increase in the optical depth at low frequencies with turnover as a free parameter (e.g. \citealp{gis13}). Regarding AME, we considered a standard average spectrum accounting for the superposition of different kinds of emitters (e.g. \citealp{ali09}) with overall normalization $A_{\rm AME}$ at $\sim$25~GHz as a free parameter. As for thermal dust, we considered a modified blackbody (BB) spectrum of the form $S_{\rm dust}=A_{\rm dust} \nu^{ \beta} BB(T_{\rm dust},\nu)$ with amplitude $A_{\rm dust}$, spectral index $\beta$, and temperature $T_{\rm dust}$ as free parameters. 

The resulting fitted parameters are reported in Table \ref{fit}. At frequencies below 10~GHz, we find that the overall emission from M31 is dominated by synchrotron emission with a slightly steep spectral index of -1.10$^{+0.10}_{-0.08}$, and amplitude at 1~GHz of 6.97$^{+0.52}_{-0.55}$~Jy. Free-free emission is marginaly detected and, at the same frequency, is found to be 0.33$\pm$0.26~Jy, reaching an intensity comparable to that of synchrotron emission at $\gtrapprox$10~GHz (i.e. $S^{\rm ff}_{\rm 10\,GHz}/(S^{\rm ff}_{\rm 10\,GHz}+S^{\rm sync}_{\rm 10\,GHz})$=0.32$\pm$0.26). The comparison between synchrotron and free-free emission at $\simeq$10~GHz and the relative spectra are discussed in the literature (e.g. \citealp{pee11,tab13}), giving results that are consistent with ours. AME shows a similar intensity to that of synchrotron and free-free emission at $\simeq$10~GHz, overtaking between 20~GHz and 50~GHz. At $\simeq$25~GHz AME shows a flux density of 1.45$^{+0.17}_{-0.19}$~Jy. Thermal dust emission dominates the emission budget at frequencies above 60~GHz, as expected. The single modified BB spectrum that we used to fit thermal dust emission shows a spectral index of 1.490$^{+0.057}_{-0.055}$ and a dust temperature of 18.80$^{+0.55}_{-0.53}$~K. The reported fit has a $\chi^{2}=6.5$ for 18 degrees of freedom (DoF). We repeated the fit by assuming only classical emission mechanisms: not accounting for an extra emission consistent with AME makes the fit clearly unnatural with an evident discrepancy between fit and data and a $\chi^{2}=162$ for 19 DoF. This is confirmed by applying the Bayesian information criterion \citep{sch78} and the Akaike information criterion \citep{aka74} resulting in a very strong indication of the necessity in the fit for the presence of an AME-like component.

In order to double-check the consistency relating to the convolution at 51$'$.3, we repeated the flux density extraction and the fit by removing the lowest-frequency $WMAP$ channel as well as the GAU$_{34.5}$, Haslam, and Reich data, and convolving all our maps to the angular resolution of the 33~GHz channel of $WMAP$ (i.e. 39$'$.1). Similarly, we repeated the flux density extraction by convolving all our maps at the resolution of the 28.4~GHz $Planck$ channel (33$'$.1) and also removing the 33~GHz channel of $WMAP$. When convolving at these finer angular resolutions, the data at frequencies below the SRT $C$-band cannot be used because they are characterized by a coarser angular resolution, with the exception of the SRT HI data. Both fits give consistent results in terms of the amount of AME, although they fail to lift the degeneracy between the thermal and nonthermal emission at low frequencies.

The fitted values are consistent at 95\%CL with similar values found in PLA15. The synchrotron amplitude at 1~GHz is consistent with the aforementioned flux density; similarly for the spectral index which, however, shows a slightly steeper behavior. This is consistent with what is observed in dusty galaxies (\citealp{pee11}; see also \citealp{tab17}), although steeper than what is expected for normal galaxies \citep{con92}. A detailed  resolved investigation of the correlation between AME, low-frequency spectral index, and star formation rate will be addressed elsewhere (E.S. Battistelli et al. 2019, in preparation).  The free-free emission is also in line with previously measured values. The synchrotron spectral index and the free-free fraction in M31 have been estimated by spectrum fitting in different studies \citep{hoe98,ber03,tab13}, obtaining a typical synchrotron spectral index of about $-1.0\pm$0.2 and a free-free fraction of (16$\pm$2)\% at 20~cm. 

Thermal dust shows a consistent dust temperature and slightly less steep spectral index (i.e., PLA15 find 18.2$\pm$1.0~K and $\beta$=1.62$\pm$0.11). This effect is related to the degeneracy among the significance of AME, the steepness of the thermal dust spectrum, and the thermal dust temperature. These kinds of degeneracies (similar to the synchrotron versus free-free one) do not impact the significance of the evidence of AME. In order to highlight the importance of $C$-band flux densities for the estimation of AME, we have repeated the parameter fit while ignoring the SRT $C$-band flux densities to yield 2.4~$\sigma$ evidence of AME, which isconsistent with PLA15. 

\begin{deluxetable}{lc}
\tablecaption{Fitted parameters arising from the galaxy spectrum with 1-$\sigma$ uncertainties. \label{fit}}
\tablehead{
\colhead{Parameter} & \colhead{Best fit}
}
\startdata
Synchrotron spectral index $\alpha$ & -1.10$^{+0.10}_{-0.08}$\\
Synchrotron amplitude at 1~GHz $A_{\rm sy}$ & 6.97$^{+0.52}_{-0.55}$~Jy\\
Free-free amplitude at 1~GHz $A_{\rm ff}$& 0.33$\pm$0.26~Jy\\
Average opacity turnover frequency & 48$\pm$6~MHz\\
Thermal dust temperature $T_{\rm dust}$& 18.80$^{+0.55}_{-0.53}$~K\\
Thermal dust spectral index $\beta$&1.490$^{+0.057}_{-0.055}$\\
Thermal dust amplitude at 3000~GHz &3180$\pm$230~Jy\\
AME amplitude at 25~GHz $A_{\rm AME}$&  1.45$^{+0.17}_{-0.19}$~Jy \\
$S^{\rm ff}_{\rm 10\,GHz}/(S^{\rm ff}_{\rm 10\,GHz}+S^{\rm sync}_{\rm 10\,GHz})$ & 0.32$\pm$0.26 \\
IR$_{\rm 3000\,GHz}$/AME$_{\rm 30\,GHz}$ & 2370$\pm$330 \\
$\chi^{2}$ (for 18 DoF)& 6.5 \\
\enddata
\end{deluxetable}

\section{Conclusions} \label{sec:conclusions}

We have extracted the flux density spectrum of M31 from the radio to the IR wavelength range. We compared the typical microwave emissions, finding strong synchrotron emission at low frequencies reaching flux densities comparable to free-free emission at $\gtrapprox$10~GHz. Thermal dust emission is clearly present at IR frequencies, and strong $\simeq 8\sigma$ evidence of AME is found in the 20--40~GHz range. The measured values are largely consistent with previous studies. The synchrotron spectral index is slightly steeper than that found by PLA15 although consistent with previous studies. The modified BB emission arising from thermal dust shows consistent dust temperature with respect to PLA15. The ratio between AME at 30~GHz and IR flux density at 3000~GHz is consistent with expectations arising from our Galaxy (e.g. \citealp{pla14b, hen16}; PLA15). In fact, we expect IR-to-microwave ratios of 3000:1 although with some scatter. \cite{pla11} found a ratio of $\simeq$2000:1 for the Small Magellanic Cloud. From our data, we find IR$_{\rm 3000\,GHz}$/AME$_{\rm 30\,GHz}$=2370$\pm$330. We repeated this comparison by calculating the dust radiance $R$ as in \cite{hen16}. Considering the frequency-integrated dust intensity $R$, \cite{hen16} obtained for our Galaxy a linear relation of AME$_{\rm 30\,GHz}$/$R$=6200$\pm$1200~(MJy/sr)/(W/m$^{2}$/sr). From our data we obtain  AME$_{\rm 30\,GHz}$/$R$=9100$\pm$1300~(MJy/sr)/(W/m$^{2}$/sr), which is slightly higher but consistent with expectations.

Depending on the flux densities used in the spectrum fit, some degeneracy arises between free-free and synchrotron emission, as well as within the parameters describing the thermal dust emission. Nevertheless, these degeneracies do not impact the significance of AME. The importance of this evidence strongly relies on the capability to sample low-frequency emission, for which $\simeq$7~GHz data are fundamental. This gives highly significant evidence for AME globally from the entire galaxy. Further $K$-band observations, with improved angular resolution, would be key to disentangling models and starting to study AME over an entire galaxy, outside our own one.

\acknowledgments

The SRT is funded by the Ministry of Universities and Research, the Italian Space Agency, and the Autonomous Region of Sardinia (RAS), and is operated as a National Facility by INAF. These observations could not have been made without the continuous support of the SRT observers and technical staff. We acknowledge the financial support of Sapienza Ateneo-2016 funding. We acknowledge the use of the $Planck$ Legacy Archive (http://www.esa.int/Planck) and of the Legacy Archive for Microwave Background Data Analysis (https://lambda.gsfc.nasa.gov/). This research made use of the CATS Database (https://www.sao.ru/cats/). We thank S. Righini for support in preparing the observing proposal and J. Fritz for the calibrated \textit{Herschel} maps. R.G.S., F.G., and J.A.R.M. acknowledge financial support from the Spanish Ministry of Economy and Competitiveness (project AYA2017-84185-P). A.M. thanks the RAS for financial support in the context of the research project CRP 18 (P.I. of the project: Dr. M. Burgay). We acknowledge the Scientific Editorial Service of the Instituto de Astrofisica de Canarias. We would like to thank the anonymous referee for the useful  comments which improved the Letter.




\begin{thebibliography}{}

\bibitem[Akaike(1974)]{aka74} Akaike, H. 1974 IEEE Trans. Aut. Cont. 19(6), 716 (1974)
\bibitem[Ali-Haimoud et al.(2009)]{ali09} Ali-Haimoud Y.  et al. 2009 MNRAS, 395, 1055 (2009)
\bibitem[Bennett et al. (2013)]{ben13} Bennett, C.L., et al. 2013, ApJS, 208, 2, 20 (2013)
\bibitem[Battistelli et al.(2006)]{bat06} Battistelli E.S. et al. 2006, ApJL, 645, L141 (2006)
\bibitem[Battistelli et al.(2015)]{bat15} Battistelli E.S. et al. 2015, ApJ, 801, 111 (2015)
\bibitem[Berkhuijsen et al.(2003)]{ber03} Berkhuijsen E.M. et al. 2003, A\&A, 398, 937 (2003)
\bibitem[BICEP2/Keck and Planck Collaborations(2015)]{bic15} BICEP2/Keck and Planck Collaborations, 2015, PhRvL 114, 101301 (2015)
\bibitem[Bolli et al.(2015)]{bol15} Bolli, P., et al., 2015, JAI, 4, 3, 1550008 (2015)
\bibitem[Condon (1974)]{con74} Condon J. J., 1974, ApJ, 188, 279 (1974)
\bibitem[Condon (1992)]{con92} Condon J. J., 1992, ARA\&A, 30, 575 (1992)
\bibitem[Condon (1998)]{con98} Condon J. J., 1998, AJ, 115, 5, 1693-1716 (1998)
\bibitem[Cruciani et al.(2016)]{cru16} Cruciani, A. et al., 2016, MNRAS, 459, 4, 4224 (2016)
\bibitem[De Zotti et al.(2018)]{dez18} De Zotti, G., et. al., 2018, JCAP, 4, 20 (2018)
\bibitem[Dickinson et al.(2018)]{dick18} Dickinson, C., et. al., 2018, NewAR, 80, 1-28 (2018)
\bibitem[Draine et al.(1998)]{dra98} Draine B.T. et al. 1998, ApJ, 494, L19 (1998)
\bibitem[Draine (2011)]{dra11} Draine B.T. 2011, Physics of the Interstellar and Intergalactic Medium (Princeton University Press) (2011)
\bibitem[Draine et al.(2012)]{dra12} Draine B.T. et al. 2012, ApJ, 757, 1 (2012)
\bibitem[Dwarakanath et al.(1990)]{dwa90} Dwarakanath, K.S.,et al., 1990, JApA. 11, 323-410 (1990)
\bibitem[Emerson et al.(1988)]{eme88} Emerson D.T et al. 1988, A\&A,190, 353 (1988)
\bibitem[Foreman-Mackey et al.(2013)]{formac13} Foreman-Mackey D. et al. 2013, PASP, 125, 306 (2013)
\bibitem[Fritz et al.(2012)]{fri12} Fritz J. et al. 2012, A\&A, 546, A34 (2012)
\bibitem[Galvin et al.(2012)]{gal12} Galvin T.J. et al. 2012, SerAJ, 184, 41 (2012)
\bibitem[Genova-Santos et al.(2015)]{gen15} Genova-Santos R. et al. 2015, MNRAS, 452, 4169 (2015)
\bibitem[Ghisellini (2013)]{gis13} Ghisellini G. 2013, Radiative Processes in High Energy Astrophysics (Springer International Publishing Switzerland) (2013)
\bibitem[Goodman et al.(2010)]{goo10} Goodman J. et al., Comm. App. Math. and Comp. Sci., 25, 1 (2010)
\bibitem[Greaves et al.(2018)]{gre18} Greaves J.S. et al., NatAs, 2, 662-667, (2018)
\bibitem[Haslam et al.(1981)]{has81} Haslam, C.G.T. et al. 1981, A\&A, 100, 209 (1981)
\bibitem[Hensley et al.(2015)]{hen15} Hensley B.S. et al. 2015, MNRAS, 449, 809 (2015)
\bibitem[Hensley et al.(2016)]{hen16} Hensley B.S. et al. 2016, ApJ, 827, 45 (2016)
\bibitem[Hensley et al.(2017)]{hen17} Hensley B.S. et al. 2017, ApJ, 836, 2, 179 (2017)
\bibitem[Hoernes et al.(1998)]{hoe98} Hoernes P. et al. 1998, A\&A, 334, 57 (1998)
\bibitem[Kogut et al.(1996)]{kog96} Kogut A. et al. 1996, ApJ, 460, 1 (1996)
\bibitem[Melis et al.(2018)]{mel18} Melis A. et al. 2018, JAI, 7, 1, 1850004 (2018)
\bibitem[Murgia et al.(2016)]{mur16} Murgia M. et al. 2016, MNRAS, 461, 4, 3516 (2016)
\bibitem[Murphy et al.(2010)]{mur10} Murphy E.J. et al. 2010, ApJL, 709, L108 (2010)
\bibitem[Murphy et al.(2018)]{mur18} Murphy E.J. et al. 2018, ApJ, 862, 1, 20, 7 (2018)
\bibitem[Paladini et al.(2015)]{pal15} Paladini R. et al. 2015, ApJ, 813, 24 (2015)
\bibitem[Papoulis et al.(1975)]{pap75} Papoulis A. et al. 1975, IEEE, 22, 735-742 (1975)
\bibitem[Peel et al.(2011)]{pee11} Peel, M.W. et al. 2011, MNRAS, 416, 1, 99 (2011)
\bibitem[Perley et al.(2013)]{per13} Perley, R.A. et al., 2013, ApJS, 206, 2, 16 (2013)
\bibitem[Planck Collaboration(2011)]{pla11} Planck Collaboration 2011, A\&A 536, A17 (2011)
\bibitem[Planck Collaboration(2014a)]{pla14a} Planck Collaboration 2014, A\&A 571, A6 (2014)
\bibitem[Planck Collaboration(2014b)]{pla14b} Planck Collaboration 2014, A\&A 565, A103 (2014)
\bibitem[Planck Collaboration(2015)]{pla15} Planck Collaboration 2015, A\&A 582, A28 (2015)
\bibitem[Prandoni et al.(2017)]{pra17} Prandoni I., et al., 2017 A\&A, 608, 40, 26 (2017)
\bibitem[Reich (1982)]{rei82} Reich, W., 1982, A\&AS, 48, 219 (1982)
\bibitem[Scaife(2010)]{sca10} Scaife, A. 2010, MNRAS, 406, 1, L45-L49 (2010)
\bibitem[Schwarz(1978)]{sch78} Schwarz, G. 1978, AnSta 6, 461 (1978)
\bibitem[Tabatabaei et al.(2013)]{tab13} Tabatabaei, F.S. et al., 2013, A\&A, 557, A129 (2013)
\bibitem[Tabatabaei et al.(2017)]{tab17} Tabatabaei, F.S. et al., 2017, ApJ, 836, 185 (2017)
\bibitem[Verkhodanov et al.(1997)]{ver97} Verkhodanov O.V., et al. 1997, Baltic Astronomy, 6, 2, 275-278 (1997)
\bibitem[Watson et al.(2005)]{wat05} Watson R.A., et al. 2005, ApJL, 624, L89 (2005)

\end{thebibliography}
\end{document}